# Pattern Synthesizing of the Concentric Ring Arrays Using Recursive Least Square Method

Atefe Akbari-Bardaskan
Department of Electrical Engineering, Ferdowsi University of Mashhad, Mashhad, Iran

*Abstract*— **This letter presents a recursive technique to synthesize the array factor (AF) of a concentric ring array. In this method, first, the problem is modeled using the traditional least square method (LSM). In the second step, a recursive technique is applied to the defined problem. It is shown that for many practical ring arrays, only a few iteration number is sufficient to obtain a solution with high accuracy. To evaluate the performance and benefits of the proposed method, various concentric ring arrays with different array factor, including equi-ripple pattern, pattern with deep nulls, flat-top array factor, are examined. It is shown that the introduced procedure is a very noble candidate for this purpose.**

*Index Terms*—Array Factor, Concentric Ring Array, Recursive Least Square Method.

## I. INTRODUCTION

In electrical engineering, especially in the communication field, single element antenna is widely utilized for numerous applications [1-6]. A single element antenna is simple, cheap, and low profile. Although a single element antenna it is a very good choice, especially for the industrial projects, but its gain is typically low. To increase the gain, several single element antennas should be arranged in a linear, planar, or ring geometry. This arrangement is called the antenna array. An array of single element antennas are extensively used in the various applications such as 5G/6G networks, seekers, and electromagnetic compatibility instruments [7-10].

All of the antenna arrays have high gain. Also, the radiation pattern of them is easily controlled by changing the magnitude and phase of excitation coefficients of each elements. To control the radiation pattern of an array, a beam-forming network (BFN) is necessary. Additionally, the beam-forming network can control the side lobe level (SLL) of the radiation pattern, increase or decrease the half-power beam-width (HPBW), directivity, etc. However, it is not easy, and there are several challenge for this purpose [11-14]. So far, several types of beam-forming networks, including BFN based on the substrate integrated waveguide (SIW), microstrip BFN, BFN with non-uniform profile, BFN on a perforated substrate, gap-waveguide BFN, and coplanar BFN [15-22].

The total radiation pattern of a linear, planar and ring arrays is shaped by picking the appropriate values of the magnetic and phase of array weights. Hereafter, synthesizing the array factor is a chief contest in electrical engineering, especially in antenna engineering. Nowadays, analytical, iterative, and algorithm-based tactics are announced for this problem, which some of them can moderate the number of array components [23-28].

Similar to the linear and planar arrays, concentric ring arrays are employed in several applications such as radars and satellites. Concentric ring arrays have a low volume occupation. Also, the beam-forming networks of them is simpler than the planar arrays.

So far, a smaller number of techniques are presented for synthesizing the radiation pattern of a concentric ring array. In [29] and [30], the algorithm-based approaches are hired to find the excitation coefficients of the array components of a non-uniform circular arrays. In these works, it is tried to suppress the maximum side lobe levels of the array factor. In [31], a new method is proposed for this type of array, which reduce the number of array rings by the simple try-error technique. In this work, conventional least square method, Richardson technique, and QR factorization are simultaneously used. The synthesis of the array factor of a concentric ring array is a complicated problem. To this end, evolutionary techniques such as Genetic Algorithm (GA), Particle Swarm Optimization (PSO) can also be employed. These algorithms offer a solution with high accuracy. However, the computational cost of these algorithms is very high. This can be a serious problem, especially for the large concentric arrays.

In this letter, and in the first step, the array factor of a concentric ring array is defined. In the second step, the traditional least square method is used to establish a system of linear equations. It is shown that using the LSM method in the traditional formation leads to a solution with low accuracy. So, a recursive technique have to be employed. For practical arrays, only a few iteration is sufficient to reach an acceptable accuracy. To assess the advantages of the proposed recursive method, several concentric arrays are considered. Numerical results approve that the proposed recursive techniques is a promising method. In the next section, theory of the method will be shown. In the following, numerical results will report. The last section includes some conclusions.

## II. MATHEMATICAL THEORY

Figure (1) shows the geometry of a concentric ring array. In a concentric ring array, the radius of all rings are not small. So, the array factor of it is independent of the azimuth angle in the spherical coordinate. If $r_n$, $N_r$, $I_n$ and $N_n$ represent the radius of ring $n$, the number of rings, elements weights for ring $n$ and the number of elements in ring $n$, respectively, the array factor is defined as follows [31].

$$F = I_0 + \sum_{n=1}^{N_r} I_n N_n J_0(kr_n u) \qquad (1)$$

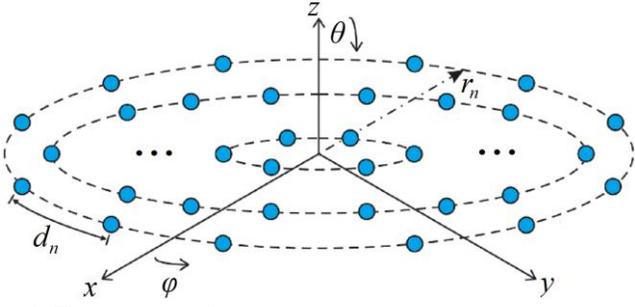

Fig. 1: The geometry of a concentric ring array.

$$u = \sin\theta \qquad (2)$$

$$N_n = \text{round}\left(\frac{2\pi r_n}{d_n}\right) \qquad (3)$$

$$d_n = 2r_n \sin(\pi/N_n) \qquad (4)$$

where $\theta$, $I_0$ and $d_n$ are the elevation angle in spherical coordinate, the weight of center element, and the arc distance between adjacent elements, respectively. It should be noted that equation (1) is used to calculate the array factor of a concentric ring array with a very good approximation. The array factor can be rewritten from its samples [32]. According to the Nyquest theorem, only $M$ samples is sufficient. So, we have.

$$F_r = \sum_{m=1}^{M} F(u_m) S(u - u_m) \qquad (5)$$

In above equation, $F_r$ and $S(.)$ are the rewritten array factor and the sampling function as follows.

$$S(u) = \frac{\sin(Mu/2)}{M \sin(u/2)} \qquad (6)$$

The smallest value of $M$ is determined as.

$$M \geq \frac{4(N_r - 1)|r_{n+1} - r_n|}{\lambda} \qquad (7)$$

where $\lambda$ is the free space wave-number. Using the sampling function, a system of linear equation is established as follows.

$$\mathbf{A}_{M\times(Nr+1)} \mathbf{I}_{(Nr+1)\times 1} = \mathbf{B}_{M\times 1} \qquad (8)$$

$$\mathbf{A} = [A_{mn}] \to A_{mn} = \begin{cases} N_n J_0(kr_n u_m) & n = 1, ..., N_r \\ 1 & n = N_r + 1 \end{cases} \qquad (9)$$

$$\mathbf{B} = [F_r(u_m)]; \quad m = 1, 2, ..., M \qquad (10)$$

$$\mathbf{I} = [I_0 \quad I_1 \quad \cdots \quad I_{Nr}]^T \qquad (11)$$

In above, $\mathbf{A}$ is the coefficient matrix of the system of equation, $\mathbf{B}$ is the vector including the array factor samples, and $\mathbf{I}$ is a column vector includes the complex excitation coefficients of array. According to the traditional least square method, the solution is found as follows.

$$\mathbf{I} = (\mathbf{A}^T \mathbf{A})^{-1} \mathbf{A}^T \mathbf{B} \qquad (12)$$

In the traditional LSM, cost of all equations is equal. So, this issue leads to the self-biasing of the problem. As a results, the accuracy of the solution will decrease. To overcome that, recursive technique is proposed.

It is assumed that in the first stage, only $M$ samples are employed to initial guess of the solution. However, in the recursive least square method, $M_0$ samples should be regarded. Our studies show that the smallest value of $M_0$ is determined as.

$$M_0 \simeq 2M \geq \frac{8(N_r - 1)|r_{n+1} - r_n|}{\lambda} \qquad (13)$$

The previous equations can be rewritten as follows.

$$\mathbf{A}_{m+1} = \begin{bmatrix} \mathbf{A}_m \\ \mathbf{a}_{m+1}^T \end{bmatrix}, \quad m = 1, 2, ..., (M_0 - M) \qquad (14)$$

$$\mathbf{a}_{m+1} = [a_{m+1}] \to$$
$$a_{m+1} = \begin{cases} N_n J_0(kr_n u_{m+M}) & n = 1, ..., N_r \\ 1 & n = N_r + 1 \end{cases} \qquad (15)$$

$$\mathbf{B}_{m+1} = [\mathbf{B} \quad \mathbf{B}_m]^T \qquad (16)$$

$$\mathbf{B}_m = F_r(u_{m+M}) \qquad (17)$$

Finally, the excitation coefficients are calculated in a recursive manner as the following equations.

$$\mathbf{I}_{m+1} = \mathbf{I}_m + \mathbf{P}_{m+1}\mathbf{a}_{m+1}\{\mathbf{B}_{m+1} - \mathbf{a}_{m+1}^T \mathbf{I}_m\} \qquad (18)$$

$$\mathbf{P}_{m+1} = \mathbf{P}_m - \mathbf{K}_{m+1}\mathbf{a}_{m+1}^T \mathbf{P}_m \qquad (19)$$

$$\mathbf{P}_m = (\mathbf{A}_m^T \mathbf{A}_m^{-1})^T \qquad (20)$$

$$\mathbf{K}_{m+1} = \mathbf{P}_m \mathbf{a}_{m+1}(\mathbf{a}_{m+1}^T \mathbf{P}_m \mathbf{a}_{m+1} + 1)^{-1} \qquad (21)$$

The studied shown that only three iterations is sufficient. To reduce the number of rings, the similar procedure shown in [31] can be considered.

### III. NUMERICAL EXAMPLES AND DISCUSSION

In the following, various concentric ring arrays with different array factor are examined. To this end, MATLAB-based program is written.

#### A. Flat-Top Array Factor

A flat-top array factor is considered as the first example. This pattern has non-zero values at $-0.4 \leq u \leq +0.4$. During the synthesis process, it is assumed $N_r=9$, $r_n=n\lambda/2$ and $N_n=\text{round}(2\pi n)$. Obtained two-dimensional array factor and the target array factor are plotted in Figure (2). It is seen that the accuracy is acceptable. The obtained ring weights are displayed in Figure (3). Also, Figure (4) shows the three-dimensional array factor. The array factor is symmetric.

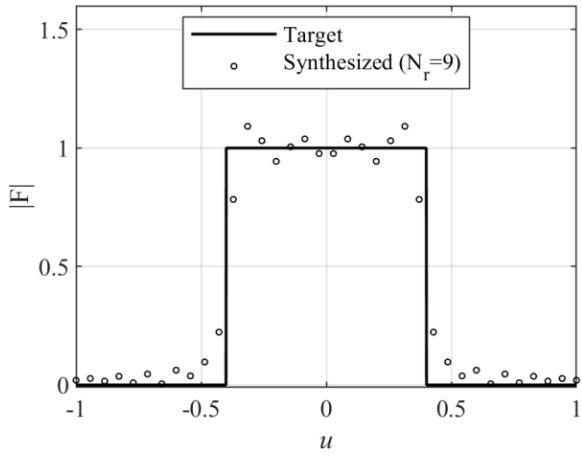

Fig. 2: The 2D array factor of the first example.

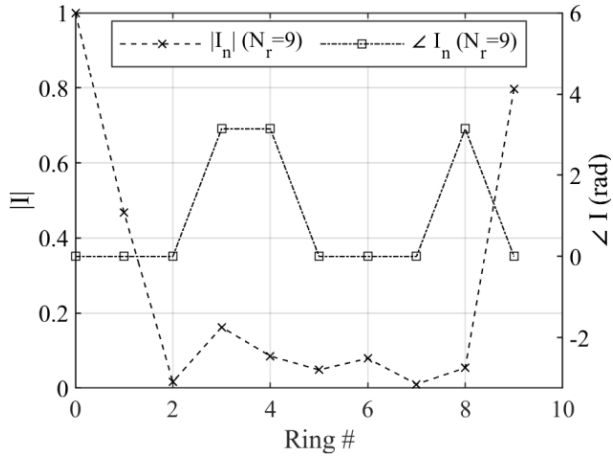

Fig. 3: The excitation coefficients of the first example.

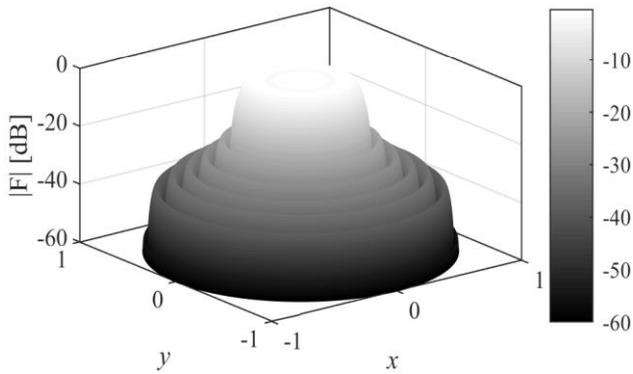

Fig. 4: The 3D array factor of the first example.

### B. Difference Array Factor

In the second example, a difference array factor with maximum side lobe level about -25 dB and $N_r$=11 is regarded. After applying the synthesis process, the obtained two-dimensional array factor and the target array factor are depicted in Figure (5). It is seen that the accuracy is reasonable. The obtained ring weights are displayed in Figure (6). Also, Figure (7) shows the three-dimensional array factor. Similar to previous example, the array factor is symmetric.

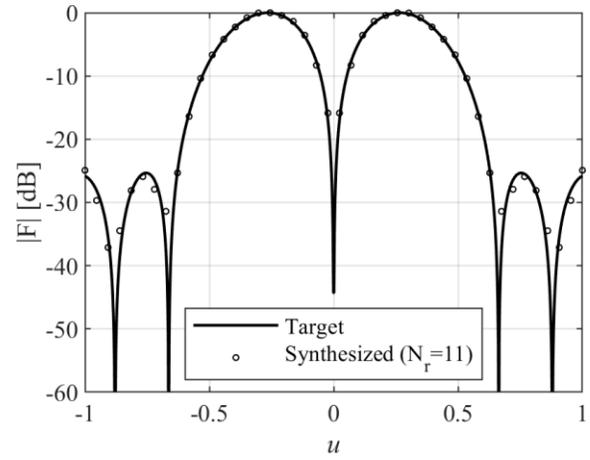

Fig. 5: The 2D array factor of the second example.

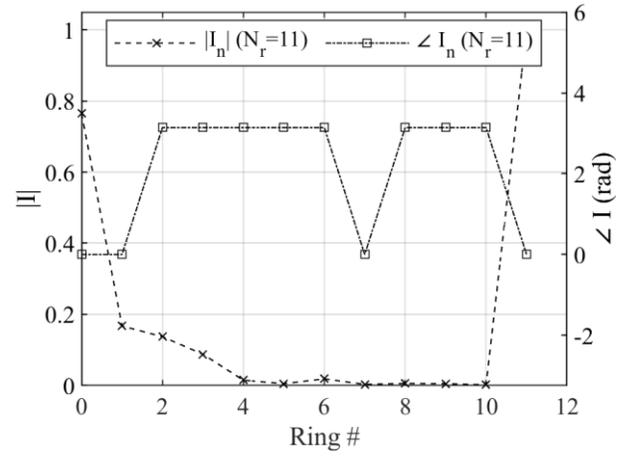

Fig. 6: The excitation coefficients of the second example.

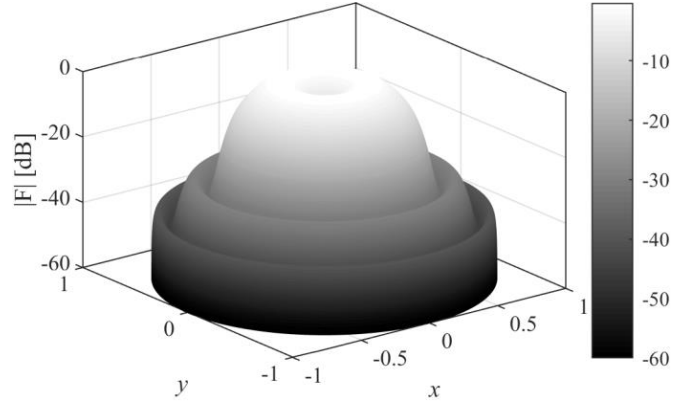

Fig. 7: The 3D array factor of the second example.

### C. Equi-Ripple Array Factor

A concentric ring array with equi-ripple array factor with side lobe level around -30 dB and $N_r$=10 is considered as the third example. The obtained two-dimensional array factor and the target array factor are depicted in Figure (8). It is seen that the accuracy is good. The obtained ring weights are displayed in Figure (9). Also, Figure (10) shows the three-dimensional array factor. Similar to previous examples, the array factor is symmetric.

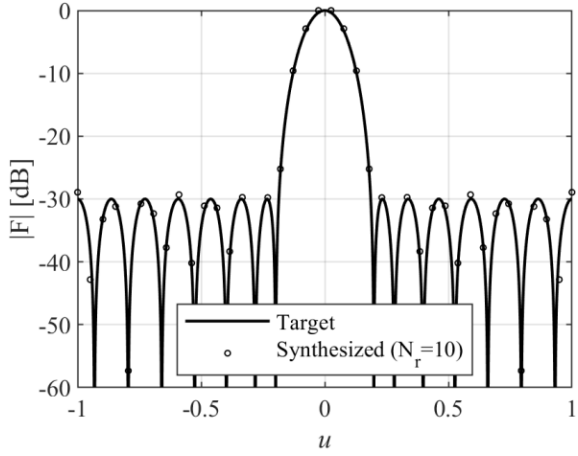

Fig. 8: The 2D array factor of the third example.

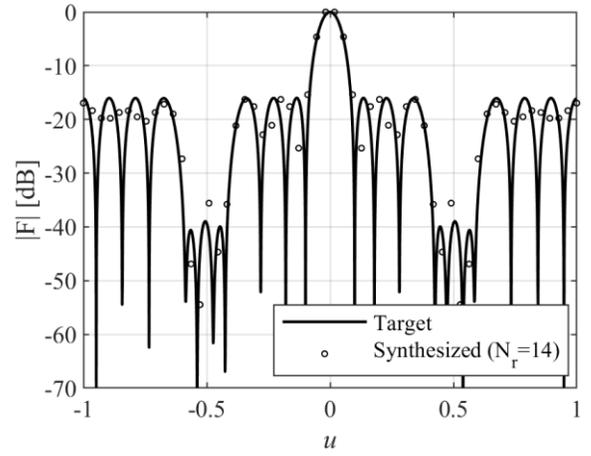

Fig. 11: The 2D array factor of the fourth example.

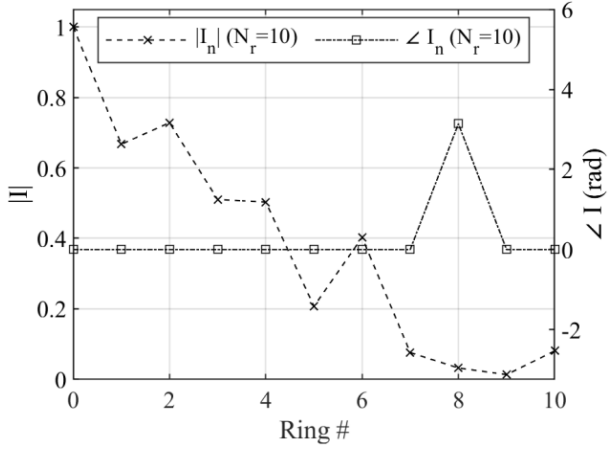

Fig. 9: The excitation coefficients of the third example.

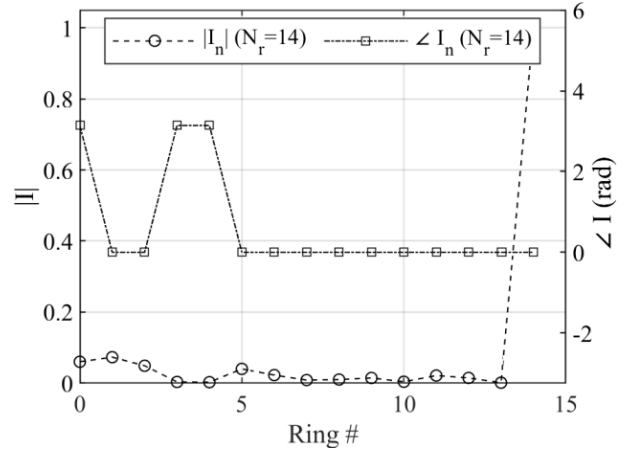

Fig. 12: The excitation coefficients of the fourth example.

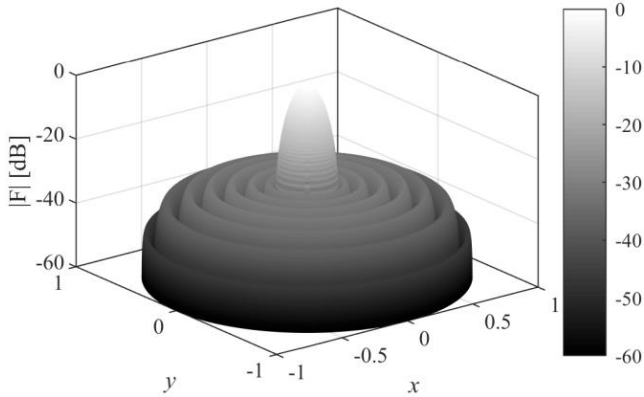

Fig. 10: The 3D array factor of the third example.

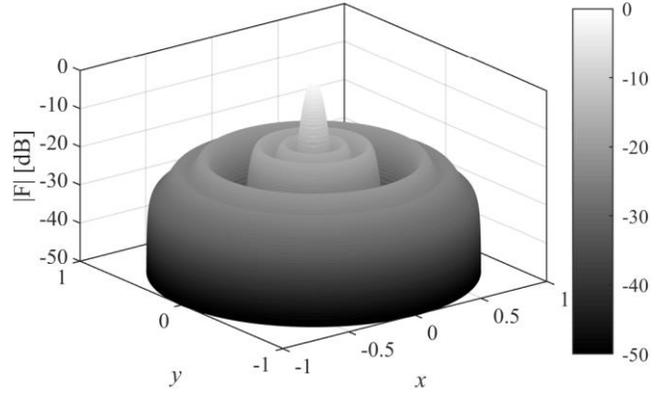

Fig. 13: The 3D array factor of the fourth example.

### D. Array Factor with Deep Nulls

An equi-ripple array factor with SLL=-16 dB, two -40 dB deep nulls, and $N_r$=14 is considered as the fourth example. The obtained two-dimensional array factor and the target array factor are depicted in Figure (11). It is seen that the accuracy is good. The obtained ring weights are displayed in Figure (12). Also, Figure (13) shows the three-dimensional array factor.

### IV. CONCLUSION

A recursive least square method is proposed to synthesize the array factor of a concentric ring array. First, the problem is modeled by the classical least square technique. Then, a recursive procedure is applied to the described problem. It is shown that for the practical concentric ring arrays, a few iteration number is enough to obtain a solution with high precision. To estimate the performance and profits of the method, several concentric ring arrays with dissimilar array factor, including equi-ripple pattern, pattern with deep nulls, flat-top array factor, are examined.